\documentclass[usenatbib,traditabstract]{aa}

\usepackage{txfonts}
\usepackage{graphicx}
\usepackage{natbib}
 
\begin{document}
 
\title{CoRoT 101186644: A transiting low-mass dense M-dwarf on an eccentric 20.7-day period orbit around a late F-star\thanks{Based on observations made with the 1-m telescope at the Wise Observatory, Israel, the Swiss 1.2-m Leonhard Euler telescope at La-Silla Observatory, Chile, the IAC-80 telescope at the Observatory del Teide, Canarias, Spain, and the 3.6-m telescope at La-Silla Observatory (ESO), Chile (program 184.C-0639).}}
 
\subtitle{Discovered in the CoRoT lightcurves}
 
\author{
L. Tal-Or \inst{1} 
\and T. Mazeh \inst{1}
\and R.\ Alonso\inst{2,3,4} 
\and F.\ Bouchy \inst{5,6} 
\and J.\ Cabrera\inst{7} 
\and H.\ J.\ Deeg\inst{3,4}  
\and M.\ Deleuil\inst{8} 
\and S.\ Faigler\inst{1}
\and M.\ Fridlund\inst{9,10} 
\and G.\ H{\'e}brard\inst{5,6} 
\and C.\ Moutou\inst{8} 
\and A. Santerne \inst{8} 
\and  B.\ Tingley\inst{3,4}
}
 
\institute{School of Physics and Astronomy, Raymond and Beverly Sackler Faculty of
Exact Sciences, Tel Aviv University, Tel Aviv, Israel\\
\email{levtalo@post.tau.ac.il}
\and
Observatoire de l'Universit\'{e} de Gen\`{e}ve, 51 chemin des Maillettes, 1290 Sauverny, Switzerland 
\and
Instituto de Astrof\'{\i}sica de Canarias, 38205 La Laguna, Tenerife, Spain
\and
Universidad de La Laguna, Dept. de Astrof\'\i sica, 38200 La Laguna, Tenerife, Spain
\and
Institut d'Astrophysique de Paris, UMR7095 CNRS, Universit\'{e} Pierre \& Marie Curie, 98bis Bd Arago, 75014 Paris, France
\and
Observatoire de Haute Provence, CNRS/OAMP, 04870 St Michel l'Observatoire, France
\and
Institute of Planetary Research, German Aerospace Centre, Rutherfordstrasse 2, 12489 Berlin, Germany
\and
Aix Marseille Universit\'e, CNRS, LAM (Laboratoire d'Astrophysique de Marseille) UMR 7326, 13388, Marseille, France 
\and
Research and Scientific Support Department, ESTEC/ESA, Keplerlaan 1, 2200AG, Noordwijk, The Netherlands 
\and
Leiden Observatory, Leiden University, P.O. Box 9513, NL-2300 RA Leiden, The Netherlands
}
 
\date{Received ... ; accepted ...}
\abstract{We present the study of the CoRoT transiting planet candidate 101186644, also named LRc01\_E1\_4780. Analysis of the CoRoT lightcurve and the HARPS spectroscopic follow-up observations of this faint (m$_V = 16$) candidate revealed an eclipsing binary composed of a late F-type primary (T$_{\rm eff}=6090\pm200$\,K) and a low-mass, dense late M-dwarf secondary on an eccentric ($e=0.4$) orbit with a period of $\sim20.7$ days. The M-dwarf has a mass of $0.096\pm0.011$\,M$_{\odot}$, and a radius of $0.104_{-0.006}^{+0.026}$\,R$_{\odot}$, which possibly makes it the smallest and densest late M-dwarf reported so far. Unlike the claim that theoretical models predict radii that are $5\%-15\%$ smaller than measured for low-mass stars, this one seems to have a radius that is consistent and might even be below the radius predicted by theoretical models.}
\authorrunning{Tal-Or et al.}
\titlerunning{Transiting low-mass dense M-dwarf on an eccentric orbit around an F-star}
\maketitle
 
\section{Introduction}
CoRoT is the first space-mission to search for transiting extrasolar planets \citep{Baglin2009,auvergne09,Deleuil2011}. Since its launch in December 2006, $26$ transiting planets have been discovered, and many other candidates await confirmation \citep{corot21b,Grziwa2012}. The discovery process of a new transiting planet includes careful analysis of the lightcurve \citep{carpano09}, as well as photometric and spectroscopic follow-up observations \citep{deeg09,bouchy09,Santerne2011EPJWC}.
 
Lightcurves with periodic transit-like signals caused by phenomena other than a transiting planet are usually referred to as \textquoteleft false positives' or \textquoteleft false alarms' \citep[e.g.,][]{Brown2003,Santerne2012}. In the CoRoT sample, for instance, the main source of such alarms are eclipsing binary systems in various configurations \citep{almenara09}. Recently, however, such systems are becoming objects of interest in and of themselves \citep[e.g.,][]{Pont2005b,c0121}.
 
Eclipsing binaries ($=$\,EBs) with low mass ratios, like M dwarfs that eclipse A-K dwarfs, draw particular attention \citep[e.g.,][]{Bentley2009,Triaud2013}. The high-quality photometric and spectroscopic data produced for these systems by photometric surveys like HAT \citep{HAT}, WASP \citep{WASP}, CoRoT, \textit{Kepler} \citep{Borucki2010Sci}, and their follow-up programs, allow for precise determination of physical properties, such as the masses and the radii of both components. This in turn enables addressing fundamental problems, such as the M-dwarf radius problem: the claim that theoretical models predict $5\%-15\%$ smaller radii than measured for low-mass stars \citep[e.g.,][and references therein]{Torres2010,Morales2010,Kraus2011}.
 
This paper presents the study of the CoRoT transiting-planet candidate CoRoT 101186644  \citep{Cabrera2009}, also named LRc01\_E1\_4780. The CoRoT lightcurve and the HARPS spectroscopic follow-up observations have led to the conclusion that LRc01\_E1\_4780 ($=$\,C4780) is an eclipsing binary composed of a late F-type primary and a low-mass dense M-dwarf secondary star on an eccentric ($e=0.4$) orbit with a period of $\sim20.7$-days.
 
Section 2 gives some details about the star and presents the CoRoT lightcurve. Section 3 describes the ground-based follow-up observations we performed. Section 4 presents our analysis of the available data to derive the main physical parameters of the system. The astrophysics of the secondary is discussed in Section 5 within the context of the M-dwarf radius problem. Finally, Section 6 presents some more general conclusions.
 
\begin{table}
\caption{Coordinates and magnitudes of C4780 and of the main contaminator in its CoRoT-photometric mask.}
\begin{tabular}{lll}
\hline
\hline
 & \textbf{LRc01\_E1\_4780} & \\
\hline
\textbf{CoRoT ID} &\textbf{2MASS ID} & \\
$101186644$ &$19265907+0029061$ & \\
\hline
\textbf{RA (J2000)} &\textbf{Dec (J2000)} & \\
$19^{h} 26^{m} 59^{s}.08$ &$00^{\circ} 29' 06''.4 $ & \\
\hline
\textbf{Filter} &\textbf{Magnitude} &\textbf{Source} \\
\textit{B} & $17.09 \pm 0.15$ & \textit{ExoDat}$^a$ \\
\textit{V} & $16.05 \pm 0.07$ & \textit{ExoDat} \\
\textit{r'} & $15.67 \pm 0.14$ & \textit{ExoDat} \\
\textit{i'} & $14.93 \pm 0.03$ & \textit{ExoDat} \\
\textit{J} & $14.03 \pm 0.03$ & 2MASS$^b$ \\
\textit{H} & $13.63 \pm 0.02$ & 2MASS \\
\textit{Ks} & $13.56 \pm 0.03$ & 2MASS \\
\hline
 & \textbf{Main contaminator} & \\
\hline
\textbf{CoRoT ID} &\textbf{2MASS ID} & \\
$101186975$ &$19265918+0029008$ & \\
\hline
\textbf{RA (J2000)} &\textbf{Dec (J2000)} & \\
$19^{h} 26^{m} 59^{s}.21$ &$00^{\circ} 29' 01''.0 $ & \\
\hline
\textbf{Filter} &\textbf{Magnitude} &\textbf{Source} \\
\textit{V} & $19.08 \pm 0.37$ & \textit{ExoDat} \\
\textit{r'} & $18.57 \pm 0.27$ & \textit{ExoDat} \\
\hline
\hline
$^a$ \citet{ExoDatDeleuil2009} & & \\
$^b$ \citet{2mass} & & \\
\end{tabular}
\label{MAG}
\end{table}
 
\section{The CoRoT lightcurve}
 
\begin{figure}
\resizebox{\hsize}{!}
{\includegraphics{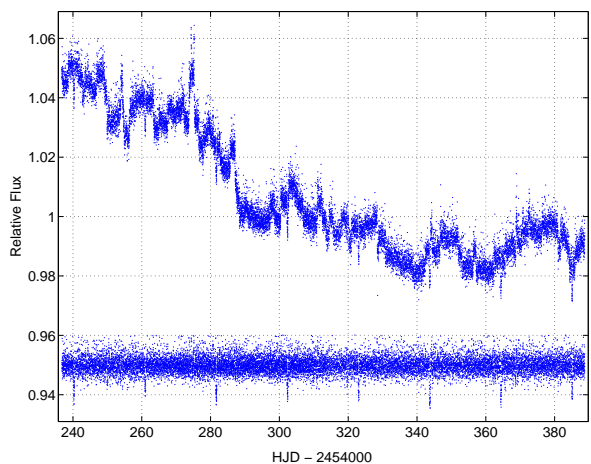}}
\caption{C4780 lightcurve normalized by its median flux value. Top: the original lightcurve. Bottom: the detrended lightcurve moved down by $0.05$, for clarity.}
\label{Fig1}
\end{figure}
 
\begin{figure}
\resizebox{\hsize}{!}
{\includegraphics{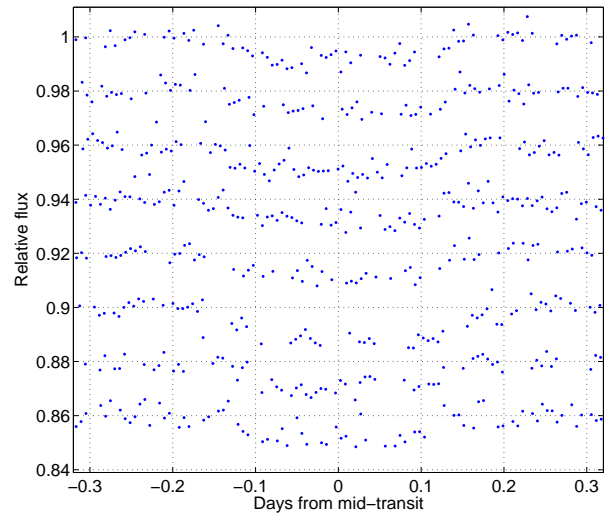}}
\caption{The $8$ transit-like events detected by CoRoT in C4780, each moved to its own timeframe (i.e. $(n\cdot P+T_0)$ days were subtracted from the time stamp of each point, where $n$ is the transit number, $T_0=2454240.3147$\,BJD, and $P=20.6837$). Subsequent events were moved down by $0.02$ for clarity.}
\label{Fig2}
\end{figure}
 
C4780 was observed by CoRoT during the LRc01 run from May 19, 2007 to Oct. 12, 2007. Table \ref{MAG} lists some basic information on its stellar properties. Since it is a relatively faint star, only monochromatic photometric data were recorded. Eight transit-like events with a depth of $\sim0.86\%$ and a period ($P$) of $20.684$ days were detected in the lightcurve \citep{Cabrera2009}. It was therefore identified as an interesting candidate for a Jovian planet residing in the so-called ``period valley'' \citep[e.g.,][]{Jones2003,Udry2003}. 
 
Figure \ref{Fig1} shows the original and detrended lightcurves of C4780, both normalized by their median flux value. Figure \ref{Fig2} zooms on the eight transit-like events detected by CoRoT in the lightcurve, each shifted to its own timeframe. Stellar and systematic variability were removed from each transit by fitting a third-degree polynomial to the out-of-transit points. 
 
\section{Ground-based follow-up observations}
As for other CoRoT transiting-planet candidates, we performed a sequence of follow-up observations to understand the true nature of C4780. Photometric observations were done to verify that the transits indeed occur on the main star inside the large-area photometric mask of CoRoT. Using spectroscopic observations we tried to distinguish it from other possible false-alarm scenarios. Together with the CoRoT lightcurve, these observations were used in determining the system's nature as an eclipsing binary system and in deriving its main physical parameters.
  
\subsection{Photometric observations}
 
\begin{figure}
\resizebox{\hsize}{!}
{\includegraphics{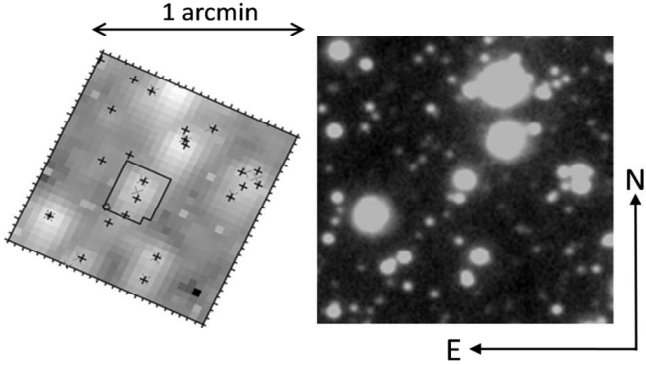}}
\caption{Images of the sky around C4780 (the star is at the center). Right: R-filter image taken on August 9, 2010, by the $1$-m Wise-Observatory telescope, Israel. Left: Image taken by CoRoT, at the same scale and orientation. The rectangular line at the center outlines the CoRoT-photometric mask. Positions of nearby stars are indicated by small crosses.}
\label{Fig3}
\end{figure}
 
Figure \ref{Fig3} shows images of the sky around C4780, taken by CoRoT and by the 1-m Wise-Observatory telescope, Israel. It can be seen that, in addition to C4780, the CoRoT-photometric mask contains at least three fainter stars, but only two of them are bright enough and fully contained in the mask to be able to cause the $\sim1\%$ transit-like signals --- the two stars $\sim6''$ south-southeast of C4780. The brighter of the two (CoRoT ID $101186975$) is fainter than C4780 by $2.9\pm0.3$\,mag in the \textit{r'}-filter (see Table \ref{MAG}), while its close neighbor is fainter than C4780 by $4.2\pm0.5$\,mag in the R-filter, as measured using the Wise-Observatory data. The contamination level inside the mask is thus $\sim10\%$.
 
ON-OFF observations of C4780, in which short timeseries during a transit and outside of it are observed and compared photometrically \citep{deeg09}, were performed during August $2010$ using the 1-m telescope at the Wise Observatory, Israel, and the Swiss 1.2-m Leonhard Euler Telescope at La-Silla Observatory, Chile. Both observations indicated that the transits occur on the main target in the mask --- C4780 -- and not on one of the two contaminators. 
 
In an effort to recover its ephemeris, C4780 was observed again on May 27, 2011, using the IAC-80 telescope at the Observatory del Teide, Canarias, Spain, and the Swiss 1.2-m Leonhard Euler telescope at La-Silla Observatory, Chile. A clear ingress was detected in these observations, which permitted the refinement of the period estimation to $20.68369\pm0.00011$\,days.
\subsection{HARPS spectra and radial velocities}
 
Spectroscopic observations of C4780 were done with the HARPS spectrograph \citep{mayor03} mounted at the 3.6-m ESO telescope, Chile, as part of the ESO large program 184.C-0639. HARPS was used with the observing mode obj\_AB, in which the background-sky spectra were recorded using the second fiber (fiber B). Overall nine spectra were recorded between June 21, 2008 and July 20, 2010 with exposure times of $45-60$\,min. The HARPS data were reduced with the online standard pipeline \citep{Baranne96,Pepe2002}. The signal-to-noise ratio ($=$\,S/N) per pixel at 550 nm is in the range of $2-6.3$, since C4780 is on the faint end in magnitude for HARPS. Radial velocities were obtained by performing weighted cross-correlation with a numerical G2 mask. The derived radial velocities ($=$\,RVs) are given in Table \ref{RVT}. 
 
\begin{table}
\caption{HARPS spectra and RVs (in km\,s$^{-1}$) of C4780.}
\begin{tabular}{ccccc}
\hline
\hline
HJD & RV & error & T$_{\rm exp}$\,(sec) & S/N\\
\hline
 $ 2454638.741570 $ & $ 16.4228 $ & $ 0.0896 $ & 3600 & $ 2.7 $\\
 $ 2454643.861050 $ & $ 14.0218 $ & $ 0.1637 $ & 3600 & $ 3.1 $\\
 $ 2455304.897570 $ & $ 14.3876 $ & $ 0.0921 $ & 3600 & $ 4.7 $\\
 $ 2455338.784230 $ & $ 18.9455 $ & $ 0.1095 $ & 3600 & $ 3.4 $\\
 $ 2455351.719640 $ & $ 27.1110 $ & $ 0.1004 $ & 3600 & $ 3.8 $\\
 $ 2455352.737020 $ & $ 28.0719 $ & $ 0.0737 $ & 3600 & $ 4.4 $\\
 $ 2455353.891020 $ & $ 26.5536 $ & $ 0.0510 $ & 3600 & $ 6.3 $\\
 $ 2455389.722540 $ & $ 15.6913 $ & $ 0.0853 $ & 2700 & $ 3.9 $\\
 $ 2455397.769130 $ & $ 22.2351 $ & $ 0.1971 $ & 2700 & $ 2.0 $\\
\hline
\hline
\end{tabular}
\label{RVT}
\end{table}
 
The spectra were also analyzed with TODMOR \citep[e.g.,][]{zm94, TODMOR, TODMOR2}, a two-dimensional correlation algorithm customized for detecting faint-secondary companion in a spectrum. However, the signature of neither the secondary in C4780 ($=$\,C4780\,B) nor any other stellar contaminant were detected.
 
To determine the fundamental atmospheric parameters of the primary in C4780 ($=$\,C4780\,A), we used the method described by \citet{Bruntt2010}, which has become the standard method for characterizing CoRoT targets since CoRoT-3b \citep[e.g.,][]{CoRoT3b,CoRoT6b}. In this method the observed spectra are first co-added to create a single master spectrum, and then a synthesized spectrum is fitted to this master spectrum using either the Spectroscopy Made Easy \citep[SME,][]{Valenti1996,VF2005} or the VWA \citep{Bruntt2008} spectral analysis packages.
 
\begin{figure}
\resizebox{\hsize}{!}
{\includegraphics{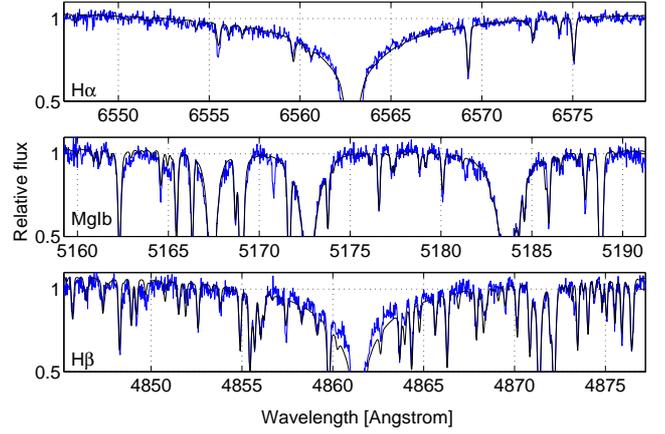}}
\caption{Three parts of the co-added HARPS spectrum of C4780 (blue), together with the fitted model spectrum (black). Top, middle, and bottom panels show $32\,\AA{}$ long parts of the spectrum, focused on the H\,$\alpha$, Mg\,I\,b, and H\,$\beta$ lines, respectively.}
\label{Fig4}
\end{figure}
 
\begin{table}
\caption{Parameters from ground-based observations.}
\begin{tabular}{lr}
\hline
\hline
Parameter & Value \\
\hline
Photometric follow-up & \\
\hline
Orbital period ($P$) [day]& $20.68369\pm0.00011$\\
Time of center of primary transit ($ T_0 $) [BJD]& $2455708.867\pm0.008$ \\
\hline
Spectrum modeling & \\
\hline
Effective temperature ($T_{\rm eff}$) [K]& $6090\pm200$\\
Surface gravity (log\,$g$) [cgs]& $4.4\pm0.2$\\
Metallicity ($[\rm{Fe/H}]$) [dex]& $+0.2\pm0.2$\\
Projected rotational velocity ($v\sin i$) [km\,s$^{-1}$]& $3\pm2$\\
Mass (M$_1$) [M$_{\odot}$]& $1.2\pm0.2$\\
Age$^*$ [Gyr]& $<7$\\
\hline
Broad-band photometry modeling & \\
\hline
Effective temperature ($T_{\rm eff}$) [K]& $5800\pm400$\\
Surface gravity (log\,$g$) [cgs]& $4.45_{-0.18}$\,$^{+0.09}$\\
Distance (d) [pc]& $1100_{-150}$\,$^{+300}$\\
Extinction (A$_V$) [mag]& $1.1_{-0.4}$\,$^{+0.2}$\\
\hline
\hline
$^*$\,$2$\,$\sigma$ upper limit& \\
\end{tabular}
\label{SPAR}
\end{table}
 
The atmospheric parameters found this way are listed in Table \ref{SPAR}. The relatively large uncertainties are a result of the low S/N of the spectra and the addition of possible systematic errors \citep[see e.g.,][]{Bruntt2010a,Bruntt2012,Torres2012}. Figure \ref{Fig4} shows three parts of the master spectrum of C4780, $32\,\AA{}$ long each, focused on the H\,$\alpha$, H\,$\beta$, and Mg\,I\,b lines, together with the fitted model spectrum. The extended wings of Balmer lines can be used to constrain $T_{\rm eff}$, while log\,$g$ of late-type stars can be determined from pressure-sensitive lines like Mg\,I\,b \citep{Bruntt2010}.
 
\section{System parameters}
 
\subsection{Isochrone fitting}
 
\begin{figure}
\resizebox{\hsize}{!}
{\includegraphics{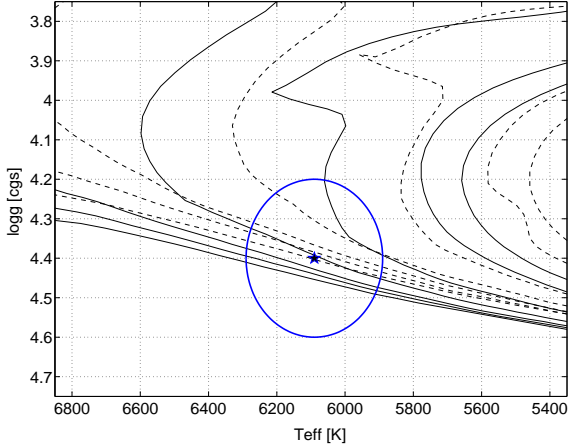}}
\caption{Two sets of Y$^2$ stellar isochrones from \citet{Demarque2004}, one for $[$Fe/H$]=0.05$ (solid lines) and one for $[$Fe/H$]=0.39$ (dashed lines). Both sets are for $[\alpha$/Fe$]=0$ and ages of $0.2$,\, $0.4$,\, $1$,\, $2$,\, $4$,\, $8$,\, and $10$ Gyr (going from left to right along the log\,$g=4.3$ line). The estimated $T_{\rm eff}$ and log\,$g$ of C4780\,A are marked by a star and their uncertainties are marked by an ellipse.}
\label{Fig5}
\end{figure}
 
The primary mass and the system's age were estimated using the atmospheric parameters derived from the HARPS spectra and a grid of Y$^2$ stellar isochrones \citep{Yi2001,Demarque2004}. This was done by taking into account all age and mass values that fall into the ellipsoid in the ($T_{\rm eff}$, log\,$g$, $[$Fe/H$]$) space defined by the atmospheric parameters and their errors. To illustrate the process Figure \ref{Fig5} shows two sets of Y$^2$ stellar isochrones of $0.2-10$\,Gyr, one for $[$Fe/H$]=0.05$ and one for $[$Fe/H$]=0.39$. The ellipse defined by the estimated $T_{\rm eff}$, log\,$g$, and their uncertainties is also shown. A lower limit of $0.2$\,Gyr on the age was set to ignore possible pre-main sequence solutions. This procedure yielded a mass estimate of $1.21\pm0.10$\,M$_{\odot}$ and an upper limit of $\sim7$\,Gyr for the system's age (at a $2$\,$\sigma$ confidence level). Following \citet{Basu2012} we have conservatively doubled the mass errors to take possible uncertainties in stellar model parameters into account.
 
The consistency between the atmospheric parameters derived from the HARPS spectra and the broad-band photometry (listed in Table \ref{MAG}) was also checked. This was done by fitting the distance and extinction to minimize the $\chi^{2}$ between each isochrone point, which also contains predictions for the true M$_V$ and color values, and the observed magnitudes. \textit{ExoDat} and 2MASS magnitudes were translated to the systems used in the isochrones using the relations given by \citet{ExoDatDeleuil2009} and \citet{Carpenter2001}. The average extinction law (R$_V=3.1$) of \citet{SM1979} were assumed, together with the \citet*{CCM1989} total-to-selective extinction ratios. The parameters were derived taking all model-points with $\chi^{2}\leq\chi^{2}_{min}+1$ into account. Only isochrones with metallicity in the range allowed by the spectrum modeling ($+0.2\pm0.2$) were considered. The results, given at the bottom of Table \ref{SPAR}, are consistent with the parameters derived from the HARPS spectra.
 
\begin{table*}
\caption{The parameters of C4780 from the combined lightcurve and RV modeling.}
\centering
\begin{tabular}{llrl}
\hline
\hline
Symbol & Parameter name & Value & Units\\
\hline
& Orbital parameters & & \\
\hline
$ P $ & Orbital period & $ 20.6841 \pm 0.0006 $ & day \\
$ e\cos\omega $ & Eccentricity $\times$ cosine longitude of periastron & $ 0.251_{-0.007}$\,$^{+0.002} $ & -- \\
$ e\sin\omega $ & Eccentricity $\times$ sine longitude of periastron & $ -0.314 \pm 0.006 $ & -- \\
$ K $ & RV semiamplitude & $ 6.816 \pm 0.039 $ & km\,s$^{-1}$ \\
$ \gamma $ & Center-of-mass RV & $ 19.608 \pm 0.038 $ & km\,s$^{-1}$ \\
$ T $ & Time of periastron$^*$ & $ 2454234.31_{-0.09}$\,$^{+0.04} $ & BJD \\
$ f $ & Mass function$^*$ & $ 0.000522 \pm 0.000010 $ & M$_{\odot}$ \\
\hline
& Photometric parameters & & \\
\hline
$ T_0 $ & Time of center of primary transit & $ 2454240.3144 \pm 0.0023 $ & BJD \\
$ J_s $ & Surface-brightness ratio & $ 0.02_{-0.02}$\,$^{+0.04} $ & -- \\
$ r_t $ & Fractional sum of radii$^{**}$ ($=($R$_1+$R$_2)/a$) & $ 0.0336_{-0.0004}$\,$^{+0.0019} $  & -- \\
$ k $ & Ratio of radii$^{**}$ ($=$R$_2/$R$_1$) & $ 0.095_{-0.001}$\,$^{+0.026} $ & -- \\
$ x $ & Impact parameter$^{**}$ ($=\cos i\cdot (1-e^2)\cdot r_t^{-1}\cdot (1+e\sin\omega)^{-1} $) & $ 0.0 \pm 0.2 $ & -- \\
$ L_3 $ & Third-light (blending) fraction$^{***}$ & $ 0.10_{-0.03}$\,$^{+0.35} $ & -- \\
$ u_p $ & Limb-darkening coefficient of primary$^{***}$ & $ 0.57_{-0.01}$\,$^{+0.06} $ & -- \\
$ i $ & inclination$^*$ & $ 90.0 \pm 0.4 $ & degree \\
\hline
& Parameters estimated assuming M$_1=1.2 \pm 0.2$\,M$_{\odot}$  & & \\
\hline
$ q $ & Mass ratio$^*$ (M$_2$/M$_1$)& $ 0.080 \pm 0.005 $ & -- \\
$ a $ & Semi-major axis$^*$ & $ 0.16 \pm 0.01 $ & AU \\
M$_2$ & Mass of secondary$^*$ & $ 0.096 \pm 0.011 $ & M$_{\odot}$ \\
R$_1$ & Radius of primary$^*$ & $ 1.07 \pm 0.07 $ & R$_{\odot}$ \\
R$_2$ & Radius of secondary$^*$ & $ 0.104_{-0.006}$\,$^{+0.026} $ & R$_{\odot}$ \\
log\,$g_1$ & Surface gravity of primary$^*$ & $ 4.47_{-0.06}$\,$^{+0.03} $ & cgs \\
\hline
\hline
 & $^*$ derived analytically from the fitted parameters & & \\
 & $^{**}$ see \citet{EBAS} and \citet{EBAS2} for the reasoning behind this definition & & \\
 & $^{***}$ lower limit was determined by the chosen prior (see text) & & \\
\end{tabular}
\label{PAR}
\end{table*}
 
\subsection{MCMC analysis}
 
The rest of the system parameters were derived by modeling the CoRoT lightcurve and the HARPS RV data simultaneously, using MCMC analysis \citep[e.g.,][App. A]{Tegmark}. The MCMC run consisted of $10^6$ accepted steps. The input data were the eight-transit data shown in Figure \ref{Fig2}, seven one-day-long parts of the detrended lightcurve where the secondary eclipses might have occurred, and the nine HARPS RVs. Errors for the photometric data were derived from the scatter of the out-of-transit points, which was found to be $\sim0.25\%$.
 
The model consisted of $12$ free parameters --- $P$, $T_0$, $J_s$, $r_t$, $k$, $L_3$, $x$, $u_p$, $e\cos\omega$, $e\sin\omega$, $K$, and $\gamma$, all detailed in Table \ref{PAR}. A lower bound of $0.07$ was set on $L_3$ to fit the estimated contamination value (see Section $3.1$), and $u_p$ was constrained to be between $0.56$ and $0.66$ according to the values given in \citet{Sing2010} and the atmospheric parameters derived from the HARPS spectra. Besides these two limitations, flat prior distributions were used.
 
At each MCMC step, the lightcurve model was calculated using EBOP \citep{Popper1981}, an algorithm for analyzing eclipsing-binary lightcurves, and the RV model was calculated using our own code for solving Kepler's equation. The sizes of the MCMC Gaussian perturbations were set by a relatively short MCMC run (of $\sim10^5$ accepted steps). The $\chi^{2}$ for each trial point was the sum of the lightcurve $\chi^{2}$ and the $\chi^{2}$ of the RV data. Following the Metropolis-Hastings algorithm, trial points with lower $\chi^{2}$ were accepted, whereas trial points with higher $\chi^{2}$ were accepted only with probability of exp($-\Delta\chi^{2}/2$), assuming the observational errors to be Gaussian \citep{Ford2005AJ}.
 
Table \ref{PAR} lists the parameters estimated by the MCMC analysis. For model parameters that showed normal posterior probability distribution, the median value of the chain is mentioned as our estimation. However, for model parameters with skewed distribution (the ones with uneven confidence limits in Table \ref{PAR}) the \textit{most probable} value is mentioned as our estimation. For most of the parameters, the confidence limits were estimated as the range of values that cover the central $68.3$\% of the chain. The lower limits of $L_3$ and $u_p$, however, were determined by the chosen priors, which indicates that the ground-based observations helped in this case for placing stronger constraints on these two limits. The orbital period estimated by the MCMC analysis is consistent with the one derived from photometric follow-up observations (see Table \ref{SPAR}).
 
\begin{figure}
\resizebox{\hsize}{!}
{\includegraphics{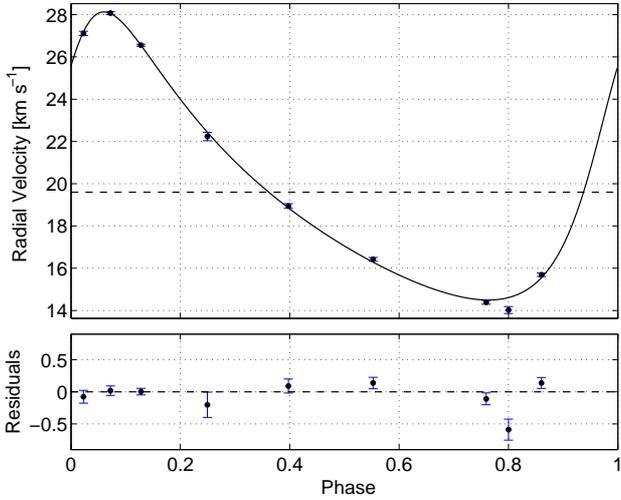}}
\caption{RVs of C4780 as measured by HARPS. The solid line is the Keplerian model produced from the orbital parameters of Table \ref{PAR}, and the dashed line is the center-of-mass velocity. The residuals are plotted in the lower panel. Note the different scale.}
\label{Fig6}
\end{figure}
 
\begin{figure}
\resizebox{\hsize}{!}
{\includegraphics{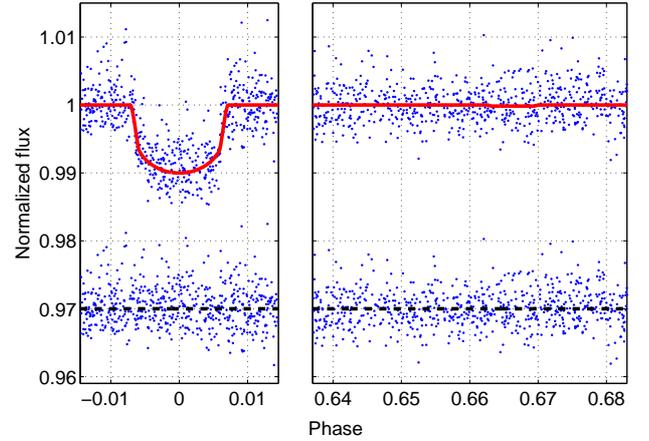}}
\caption{The primary transit (left panel) and the part where the secondary eclipse should have occurred (right panel) in the phase-folded lightcurve of C4780. Our best model is overplotted with a solid red line. The residuals were moved up by $0.97$ for clarity.}
\label{Fig7}
\end{figure}
 
Figure \ref{Fig6} presents the phase-folded RV curve of C4780 with our best orbital solution, and Figure \ref{Fig7} presents the phase-folded lightcurve and our best model. It can be seen that no secondary eclipse was detected. Given the typical noise ($\sim0.25\%$) and the expected number of points inside the secondary eclipse ($\sim100$), the maximum depth of the secondary eclipse is $\lesssim0.05\%$ ($1$\,$\sigma$ upper limit), which corresponds to a surface brightness ratio of $J_s \lesssim 0.06$.
 
\begin{figure}
\resizebox{\hsize}{!}
{\includegraphics{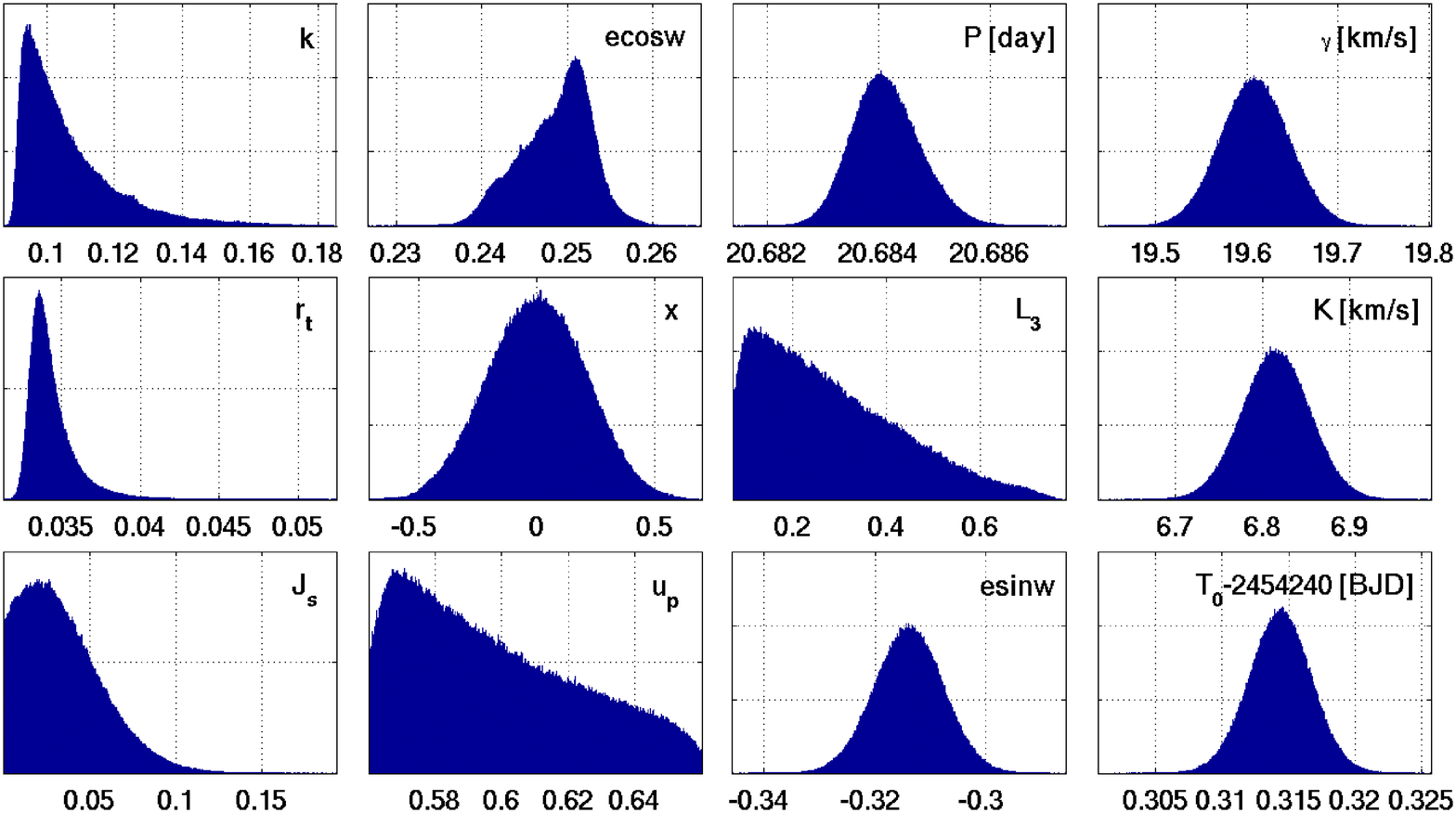}}
\caption{Histograms of the $12$ free parameters of the model from the final MCMC chain.}
\label{Fig8}
\end{figure}

\begin{figure*}
\resizebox{\hsize}{!}
{\includegraphics{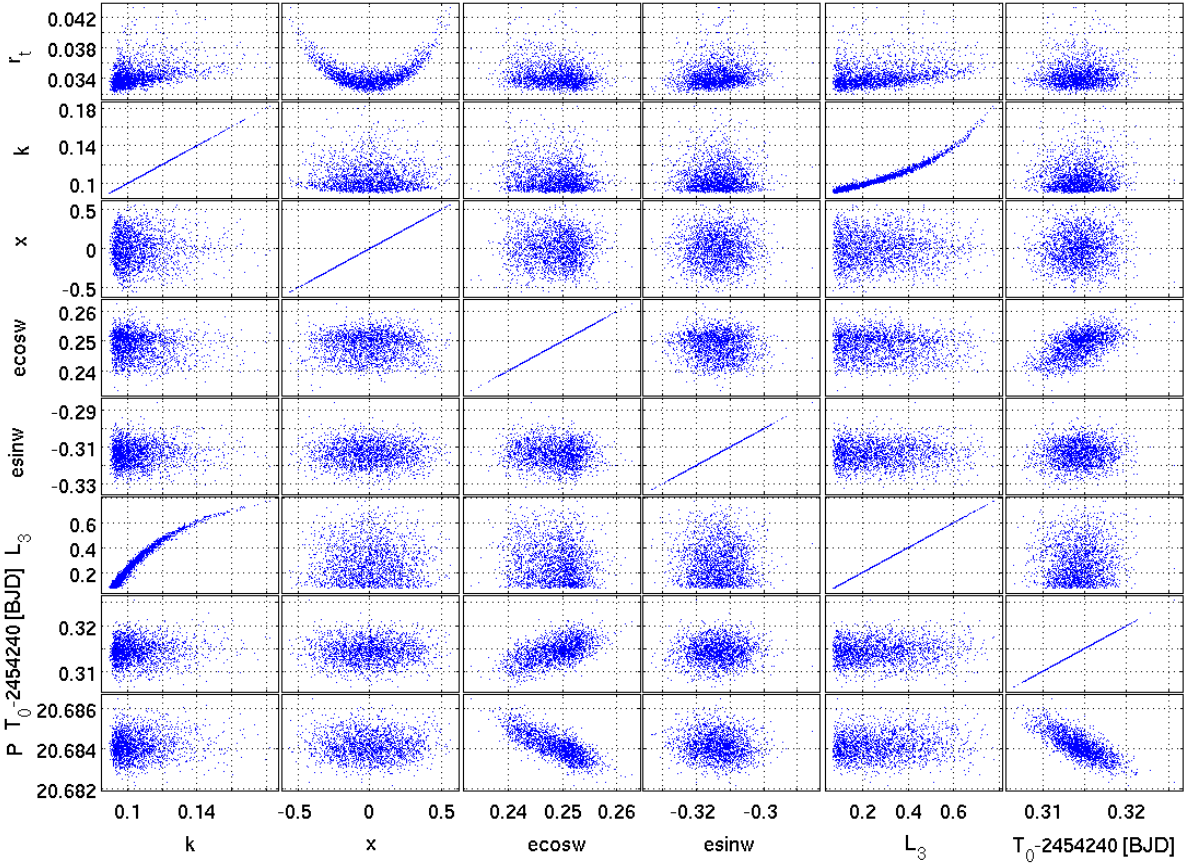}}
\caption{Selected correlation plots of parameter pairs from the final MCMC chain. For clarity, only one of every $500$ points of the chain was plotted.}
\label{Fig9}
\end{figure*}

Figure \ref{Fig8} shows the histograms of the $12$ free parameters of the model, and Figure \ref{Fig9} shows some correlation plots of parameter pairs that show non-zero correlation, both produced from the final MCMC chain. The most skewed histograms are those of $L_3$, $k$, and $r_t$. The most prominent (and obvious) correlation is between $L_3$ and $k$. These facts reflect the difficulty of setting an upper limit for $L_3$ directly from the photometric data, and as a consequence the difficulty of setting an upper limit for the secondary radius (even if the primary radius is known).
 
The secondary mass (M$_2$) and both radii (R$_1$ and R$_2$) were estimated using the MCMC chain and the estimated mass of the primary ($1.2 \pm 0.2$\,M$_{\odot}$) in the following manner: 
 
\begin{itemize}
 \item For each point of the MCMC chain the primary mass function ($f$), inclination ($i$), and semimajor axis of the primary ($a_1$) were calculated analytically.
 \item A sample of $5000$ M$_1$ values, normally distributed according the M$_1$ value found, was generated.
 \item A subsample of $5000$ points was taken from the MCMC chain by taking every $200$th point of the chain.
 \item For each M$_1$ value of the selected sample and each of the selected MCMC points, the mass ratio ($q$) was calculated analytically, using the relation
\begin{equation}
(M_1f^{-1}\sin^3i)q^3-q^2-2q-1=0 $.$
\end{equation}
The secondary mass (M$_2$) was then calculated for each pair of M$_1$ and $q$ values, and the semimajor axis ($a$) was calculated for each set of M$_1$, M$_2$, and $a_1$ values.
 \item Using Kepler's third law and the definitions of $r_t$ and $k$ (see Table \ref{PAR}) the primary radius (R$_1$) can be expressed as
\begin{equation}
R_1^3=\frac{GP^2(M_1+M_2)}{4\pi^2}\frac{r_t^3}{(1+k)^3} $ , $ 
\end{equation}
where $G$ is the universal gravitational constant and $P$ the orbital period. Using this expression, the primary and secondary radii (R$_1$ and R$_2$) were calculated analytically for each M$_1$ value of the selected sample and each of the selected MCMC points. Given the mass and radius of the primary, its surface gravity (log\,$g_1$) was also calculated.
\end{itemize}
 
The values and errors of all parameters calculated analytically ($f$, $i$, $a$, $q$, M$_2$, R$_1$, log\,$g_1$, and R$_2$) were estimated the same way as the values and errors of the free parameters of the model. These parameters are shown at the bottom part of Table \ref{PAR}. The highly uneven confidence limits of R$_2$ reflect our choice of taking the \textit{most probable} value as our estimate for model parameters with skewed distribution. The strong correlation between $k$ and $L_3$ (Figure \ref{Fig9}) shows that the estimated R$_2$ value reflects our estimation of $L_3$ to be $\sim10\%$. This contamination level agrees with the expected one, taking the contaminants inside the CoRoT-photometric mask (see Section $3.1$) into account. It means that the adopted $L_3$ value matches the nonexistence of any other, unresolved, luminous object close to C4780. This is also consistent with the nondetection of a secondary companion in the spectra with TODMOR (see Section $3.2$).
 
\section{Discussion}
 
\begin{table*}
\caption{Main properties of the stars presented in Figure \ref{Fig10}.}
\centering
\begin{tabular}{lcccccccc}
\hline
\hline
 star & mass & radius & $[\rm{Fe/H}]$ & T$_{\rm eff}$ & magnetic activity & $P_{orb}$ & $v\sin i$ & ref. \\
 name & [M$_{\odot}$] & [R$_{\odot}$] & [dex] & [K] & ($L_X/L_{bol}$) & [day] & [km\,s$^{-1}$] & \\
\hline
OGLE-TR\,123\,B$^b$     & $0.085\pm0.011$           & $0.133\pm0.009$               & ---     & ---    & ---                    & $1.8039$  & ---           & 1\\
SDSS\,0857+03\,B$^c$    & $0.090\pm0.010$           & $ 0.110\pm0.004 $             & ---     & ---    & ---                    & $0.06528$ & ---           & 2 \\
J1219-39\,B$^b$         & $0.091\pm0.002$           & $0.1174^{+0.0071}_{-0.0050}$  & $-0.21$ & ---    & ---                    & $6.7600$  & ---           & 3\\
OGLE-TR\,122\,B$^b$     & $0.092\pm0.009$           & $0.120^{+0.024}_{-0.013}$     & $+0.15$ & ---    & ---                    & $7.2687$  & ---           & 4\\
C4780\,B$^b$            & $0.096\pm0.011$           & $ 0.104_{-0.006}^{+0.026} $   & $+0.20$ & ---    & ---                    & $20.6841$ & ---           & this work \\
NN\,Ser\,B$^c$          & $0.111\pm0.004$           & $ 0.149\pm0.002 $             & ---     & ---    & ---                    & $0.13008$ & ---           & 5 \\
GK\,Vir\,B$^c$          & $0.116\pm0.003$           & $ 0.155\pm0.003 $             & ---     & ---    & ---                    & $0.34433$ & ---           & 6 \\
OGLE-TR\,106\,B$^b$     & $0.116\pm0.021$           & $0.181\pm0.013$               & ---     & ---    & ---                    & $2.5359$  & $3.59\pm0.26$ & 7\\
GJ\,551$^a$             & $0.118\pm0.012$           & $0.141\pm0.007$               & $+0.19$ & $3054$ & $2.73$E$-4\pm6.5$E$-5$ & n/a       & ---           & 8,9,10,11\\
HAT-TR-205-013\,B$^b$   & $0.124\pm0.010$           & $0.1670\pm0.0060$             & $-0.20$ & ---    & ---                    & $2.2307$  & ---           & 12\\
SDSS\,0138-00\,B$^c$    & $0.132\pm0.003$           & $ 0.165\pm0.001 $             & ---     & ---    & ---                    & $0.07276$ & ---           & 13 \\
KIC\,1571511\,B$^b$     & $0.141^{+0.005}_{-0.004}$ & $0.1783^{+0.0014}_{-0.0017}$  & $+0.37$ & ---    & ---                    & $14.0225$ & ---           & 14\\
GJ\,699$^a$             & $0.146\pm0.015$           & $0.1869\pm0.0012$             & $-0.39$ & $3222$ & $2.88$E$-6\pm3.3$E$-7$ & n/a       & $<2.8$        & 8,11,15,16\\
SDSS\,1210+33\,B$^c$    & $0.158\pm0.006$           & $ 0.200\pm0.004 $             & $-2.0$  & ---    & ---                    & $0.12449$ & ---           & 17 \\
SDSS\,1548+40\,B$^c$    & $0.173\pm0.027$           & $ 0.181\pm0.015 $             & ---     & ---    & ---                    & $0.185$   & ---           & 18 \\
RR\,Cae\,B$^c$          & $0.1825\pm0.0131$         & $ 0.2090\pm0.0143 $           & ---     & $3100$ & ---                    & $0.304$   & ---           & 19 \\
2MASS\,0446+19\,B$^b$   & $0.190\pm0.020$           & $ 0.210\pm0.010 $             & ---     & $2900$ & ---                    & $0.61879$ & ---           & 20 \\
\hline
\hline
\end{tabular}
\textbf{References:} (1) \citet{Pont2006OT123}; (2) \citet{Parsons2012a}; (3) \citet{Triaud2013}; (4) \citet{Pont2005a}; (5) \citet{Parsons2010}; (6) \citet{Parsons2012b}; (7) \citet{Pont2005b}; (8) \citet{Boyajian2012}; (9) \citet{Demory2009}; (10) \citet{Edvardsson1993}; (11) \citet{LopezMorales2007}; (12) \citet{Beatty2007}; (13) \citet{Parsons2012c}; (14) \citet{Ofir2012}; (15) \citet{Lane2001}; (16) \citet{Rojas-Ayala2012} (17) \citet{Pyrzas2012}; (18) \citet{Pyrzas2009}; (19) \citet{Maxted2007}; (20) \citet{Hebb2006}. \textbf{Notes:} $^a$ A single star for which the mass was inferred from a mass-luminosity relation. $^b$ An sb1 EB for which the mass estimation is model dependent. $^c$ An eclipsing white dwarf\,$+$\,M-dwarf binary for which the radius listed in the table is the volume averaged one.
\label{STR}
\end{table*}
 
\begin{figure}
\resizebox{\hsize}{!}
{\includegraphics{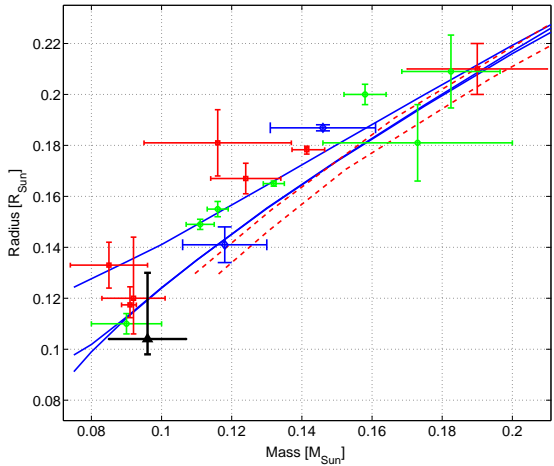}}
\caption{Mass-radius diagram of VLMS (M\,$\lesssim0.2$\,M$_{\odot}$). The symbols represent observed stars, while the lines correspond to theoretical mass-radius relations. Red rectangles are secondary stars of main-sequence EBs, green circles are secondary stars of white dwarf\,$+$\,M-dwarf EBs, and blue diamonds are single stars (all references are given in Table \ref{STR}). The black triangle stands for C4780\,B. Solid blue lines correspond to theoretical isochrones of solar metallicity and ages of $0.25,1,$ and $5$ Gyr (going from top to bottom along the $0.08$\,M$_{\odot}$ line) from \citet{Baraffe1998}. To illustrate the effect of metallicity on size dashed-red lines show the Dartmouth isochrones \citep{Dartmouth2008} of $1$ Gyr for $[\rm{Fe/H}]=0.2$ (upper line) and $[\rm{Fe/H}]=-0.5$ (lower line).}
\label{Fig10}
\end{figure}
 
Many studies of eclipsing binaries with low-mass ($0.2-0.8$\,M$_{\odot}$) main-sequence stars indicated a disagreement between theoretical models and observational data. In particular, it has been claimed that the stellar radii computed from models in this mass domain are $5\%-15\%$ lower than observed
\citep[e.g.,][]{LopezMoralesRibas2005,Morales2010,Kraus2011}. This argument is sometimes called \textit{the M-dwarf radius problem} \citep[e.g.,][]{Triaud2013}, and it dates back to the precise masses and radii measurements of both components of CM Draconis \citep{Lacy1977}. However, in the domain of very low-mass stars ($=$\,VLMS, M\,$\lesssim0.2$\,M$_{\odot}$), only a few studies have been published \citep[e.g.,][]{LopezMorales2007,Boyajian2012}, since only a handful of precise masses and radii have been derived in this domain. The derivation of the mass and radius of C4780\,B, albeit not a very accurate determination, is an opportunity to revisit the VLMS mass-radius relation.

In Figure \ref{Fig10} we plotted the mass and radius of C4780\,B together with other VLMS with \textit{derived} masses and radii,  the parameters of which were given in Table  \ref{STR}, together with a few available VLMS models. The figure suggested that C4780\,B might be the smallest main-sequence star detected so far\footnote{\citet{Berger2009} reported a radius measurements of the primary in the L-dwarf binary 2MASS\,0746+20 using radio emission. They derived $0.078\pm0.010$\,R$_{\odot}$ for this $0.085\pm0.010$\,M$_{\odot}$ L-dwarf. However, since this result is still under debate \citep[e.g.,][]{Konopacky2012}, we decided not to include 2MASS\,0746+20 in the sample listed in Table \ref{STR}.}. Its radius is consistent and might even be below the radius predicted by theoretical models for an M-star with such mass, metallicity, and age.

The models presented in Figure \ref{Fig10} display a theoretical spread, probably caused mainly by differences in age and metallicity. Several additional parameters, not plotted in Figure \ref{Fig10}, can have further impact on the mass-radius relation. In particular, fast rotation, magnetic activity, strong irradiation, and clouds were considered in the literature \citep[e.g.,][]{LopezMorales2007,Chabrier2007,Morales2010,corot15b,Burrows2011,Knigge2011}. Those effects can enlarge the theoretical spread of the models even further. Given this relatively large range of theoretical stellar radii, no apparent inconsistency between the observed systems plotted in Figure \ref{Fig10} and the available theory can be deduced at this point.

\section{Conclusions}
C4780\,B is a transiting very-low-mass M-dwarf, whose radius is consistent and might even be smaller than the predictions of theoretical models. Its discovery adds an important piece to the puzzle that any modern stellar-evolution theory of VLMS, both in binaries and as single stars, would have to account for. Further investigation of the blending fraction ($L_3$), for instance with a multicolor high-precision photometric observations, might reduce the uncertainties on its radius, thereby placing even stronger constraints. Since M-stars represent an overwhelming fraction of the galactic stellar population \citep[e.g.,][]{Chabrier2003,Kroupa2011arXiv}, understanding their properties is important not only for advancing stellar astrophysics but also for other fields of science, like terrestrial exoplanet searches \citep{Scalo2007}.

The size of VLMS is comparable to that of giant planets, making such objects in binary systems one of the main false-alarm sources in any transiting-planet survey \citep[e.g.,][]{Pont2006OT123,Santerne2012}. As a result, many such systems were considered as good transiting-planet candidates and were followed-up spectroscopically to determine their nature.
Complete and homogeneous analysis of the systems that were detected by CoRoT and \textit{Kepler}, similar to what \citet{Triaud2013} are aiming at using the WASP data, might considerably enhance our understanding of the astrophysics close to the low end of the main sequence.

\begin{acknowledgements}
We thank Gil Sokol for his help with the lightcurve analysis. We also thank the anonymous referee and the editor for their fruitful comments and suggestions. The CoRoT space mission, launched on 2006 December 27, was developed and is operated by the CNES, with participation of the Science Programs of ESA, ESA's RSSD, Austria, Belgium, Brazil, Germany, and Spain. Some of the data presented were acquired with the IAC80 telescope operated at Teide Observatory on the island of Tenerife by the Instituto de Astrof{\'\i}sica de Canarias. The research leading to these results has received funding from the European Research Council under the EU's Seventh Framework Program (FP7/(2007-2013)/ ERC Grant Agreement No. 291352). This research was also supported by the Israel Science Foundation (grant No.
1423/11). The IAC team acknowledges support by grant AYA2010-20982-C02-02 of the Spanish Ministerio de Econom{\'\i}a y Competividad. We also acknowledge support from the French National Research Agency (ANR-08-JCJC-0102-01).
\end{acknowledgements}

\bibliographystyle{aa}
\bibliography{C4780.bib}

\end{document}